\documentclass{ws-ijmpd}
\usepackage{graphicx}
\usepackage{dcolumn}
\usepackage{bm}
\usepackage{epsfig}
\usepackage{epsfig}
\input psfig.sty

\def\la{\mathrel{\hbox{\rlap{\hbox{\lower4pt\hbox{$\sim$}}}\hbox{$<$}}}}

\begin{document}
\title{Constraints on the  Hubble Parameter from galaxy clusters and the Validity of the Cosmic Distance Duality Relation  }

\author{R. F. L. Holanda}

\address{Observat\'orio Nacional, 20921-400, Rio de Janeiro - RJ, Brasil}

\maketitle

\begin{history}
\received{Day Month Year}
\revised{Day Month Year}
\comby{Managing Editor}
\end{history}

\begin{abstract}
Constraints on the Hubble parameter, $H_0$, via X-ray surface brightness and  Sunyaev-Zel'dovich effect (SZE) observations of the galaxy clusters  depend on the validity of the cosmic distance duality relation (DD relation), $\eta= D_{L}(z)(1+z)^{-2}/D_{A}(z) = 1$, where $D_L$ and $D_A$ are the luminosity distance and angular diameter distance (ADD), respectively. In this work, we argue that if the DD relation does not hold the X-ray plus SZE technique  furnishes a $H^{*}_{0}=H_{0}/\eta^{2}$. We use 25 ADD of galaxy clusters to obtain simultaneous constraints on $H_{0}$ and  possible violation of the DD relation in a flat $\Lambda$CDM model. Such a  violation is parametrized  by two functions: $\eta(z) = 1 + \eta_{0}z$ and $\eta(z) = 1 + \eta_{0}z/(1+z)$, where $\eta_0$ is a constant parameter quantifying possible departures from the strict validity. Finally, by marginalizing on the $\eta_{0}$  in both parameterizations, we obtain constraints on $H_0$ regardless of the validity of the DD relation. For the linear and non linear $\eta(z)$ functions, we obtain  $H_{0}= 75^{+ 7}_{-7}$ km/s/Mpc and  $H_{0}= 75^{+ 10}_{-7}$ km/s/Mpc, respectively (without systematic erros). Our results support recent $H_{0}$ measurements by using X-ray and SZE observations of galaxy clusters  which have taken the distance duality as valid. 
\end{abstract}

\keywords{Hubble parameter, distance cosmic,galaxy clusters,Sunyaev-Zel'dovich effect}

\section{Introduction}\label{sec:introduction}


An important phenomenon occurring in hot gas of the galaxy clusters is the Sunyaev-Zel'dovich Effect (SZE) \cite{SunZel72,itoh}, a small distortion on the Cosmic Microwave Background (CMB) spectrum provoked by the inverse Compton scattering of the CMB photons passing through  hot plasma present in the intracluster medium. The SZE is proportional to the pressure intregrated along the line of sight, $\Delta T_{0}/T_0 \propto \int n_{e} T_{e}dl$, where { $\Delta T_{0}$ is the  decrement in the CMB temperature due to SZE},  { $T_0$ is the CMB temperature}, $n_e$ and $T_e$ are density and temperature of electrons, and its magnitude is $\Delta T_{0}/T_0 \approx 10^{-5}$ independent of galaxy cluster redshift. Another fundamental process in plasma cluster is  X-ray emission (mainly through thermal Bremsstrahlung) which  has a different dependence on eletronic density, $S_{X}\propto \int n_{e}^{2}\Lambda_{eH}$, where { $\Lambda_{eH}$ is  X-ray cooling function}. It is possible to take advantage of the different eletronic density dependencies in these phenomena and with some assumptions about  geometry of galaxy cluster evaluate its angular diameter distance (ADD), given by \cite{caval,Boname06,reese02,DeFilippis05}
\begin{equation}
D_{A}(z)\propto \frac{(\Delta T_{0})^{2}\Lambda_{eH0}}{S_{X0}T_{e}^{2}}\frac{(1+z)^{-4}}{\theta_{c}},
\end{equation}
{  where  $S_{X0}$ is  central X-ray surface brightness, $z$ is  redshift of galaxy cluster and $\theta_{c}$ is its  core radius obtained from  SZE and X-ray analysis and $\theta_c$ is the core radius.}The main advantage of this method to measure distances is that it does not rely on extragalactic distance ladder being fully independent of any local calibrator \cite{bartlett,1,2,3,4,5,6,7,8,9}. 

In the last decade some works have explored the potentiality of the SZE/X-ray technique to obtain constraints on the Hubble parameter\cite{Boname06,reese02,Kobayashi96,10,11,12,13,cml,hcl}, $H_{0}$, and also to explore the kinematic history of the universe \cite{LHC2,wang}. Likewise, various sources of uncertainty  (statistical and  systematic) in this technique, such as, shape of the cluster, isothermality, presence of clumps, X-ray and SZE calibrations, kinetic SZE, radio halos, CMB anisotropy, X-ray background, etc, have been discussed in literature. These contributions of errors added in quadrature give an error of $\approx 20$\% (statistical) and $\approx 12-15$\% (systematic) on the ADD estimated (see details in Refs. \refcite{Boname06,reese02}). 

However, the SZE/X-ray technique to measure galaxy clusters distances is strongly dependent on the validity of the cosmic distance duality  relation  (DD relation) \cite{uzan}. Such relation is the astronomical version  of the Etherington reciprocity law \cite{eth33} and relates the \textit{luminosity
distance} $D_{\scriptstyle L}$ with the \textit{angular diameter
distance} (ADD) $D_{\scriptstyle A}$ through the identity
\begin{equation}
  \frac{D_{\scriptstyle L}}{D_{\scriptstyle A}}{(1+z)}^{-2}=1.
  \label{eq0}
\end{equation}
The equality in Eq. (2) can easily be proved in the context of
Friedmann-Lema\^itre-Robertson-Walker (FLRW) cosmologies. Although taken for granted in virtually all analyses in cosmology, the verification of the observational validity of the relation is, perhaps, one of the major under-appreciated open problems in modern cosmology. This relation can be violated, for 
instance, if there are deviations from a metric theory of gravity, photons not traveling
along unique null geodesics, variations of fundamental constants, etc. 
So, any observational deviation from Eq.(2) would ignite a  crisis in observational cosmology and would be a theoretical catastrophe \cite{ellis07,ellis08}.

As commented in Ref. \refcite{uzan}, when the relation does not hold, such as 
 \begin{equation}
  \frac{D_{\scriptstyle L}}{D_{\scriptstyle A}}{(1+z)}^{-2}=\eta ,
  \label{eq0}
\end{equation}
the ADD determined of  galaxy clusters from their SZE and X-ray observations is $D^{SZE/X-ray}_{A}(z)=D_{A}(z)\eta^{2}$. Such a quantity reduces to the standard ADD only when the DD relation is strictly valid ($\eta=1$). In order to quantify the
$\eta$ parameter, some authors  fixed  $D_A(z)$  by assuming some
cosmological model. In Ref. \refcite{uzan}, the cosmic concordance model \cite{spergel03} was used to obtain the real ADD while for
$D^{SZE/X-ray}_{A}(z)$ was considered 18 ADD of galaxy clusters \cite{reese02} for which a spherically symmetric
cluster geometry was assumed. By assuming $\eta$ constant,
the statistical analysis provided $\eta = 0.91^{+ 0.04}_{-0.04}$
(1$\sigma$), and, therefore only marginally consistent with the
standard result, $\eta=1$. In this way, the authors in Ref. \refcite{bem06} also searched for deviations in the DD relation by using  38 ADD of galaxy clusters  provided by Bonamente  {\it et al.} (2006) \cite{Boname06}. They obtained  $\eta = 0.97^{+0.03}_{-0.03} $ at $1\sigma$ in the framework of the cosmic concordance $\Lambda$CDM model. Further, a deformed DD relation ($D_L=D_A(1+z)^{2+\epsilon}$) was used in Ref. \refcite{lverde,avgous} to constrain the cosmic opacity by combining a recent type Ia supernovae (SNe Ia) compilation \cite{union} with the latest measurements of the Hubble expansion at redshifts between 0 and 2 \cite{lverde2}. They found $\epsilon=-0.04_{-0.07}^{+0.08}$ (2-$\sigma$).  However, what it was really done in these works was a consistency test
between the assumed cosmological model and some results provided by a
chosen set of astrophysical phenomena. 
  
In recent papers, the validity of the DD relation has been discussed in a model-independent cosmological
test by using ADD of  galaxy clusters and luminosity distances of the SNe Ia \cite{Holapjl,Hola2011,Deepak,Hongwei2011}. Basically, the SNe Ia redshifts  were carefully chosen to coincide with the ones of the
associated galaxy cluster sample ($\Delta z<0.005$),  allowing
a direct test of DD relation. However, the galaxy clusters and SNe Ia observations has different systematics error sources and the influence of the SNe Ia light curve fitters on the test, for instance, has been discussed in Ref. \refcite{Hola2011}.
   

In this work, by assuming that a more general expression as Eq. (3) is valid, we initially explore the sensitivity in the $H_{0}$ determination from SZE/X-ray technique on the $\eta(z)$ value. Constraints on $H_{0}$ and possibles deviations of DD relation are obtained simultaneously  in a flat $\Lambda$CDM model  by using 25 ADD of galaxy clusters of the De Filippis {\it et al.} sample\cite{DeFilippis05}. The mass density parameter ($\Omega_{m}$) is constrained by using a joint analysis involving the ADD from clusters and the baryon acoustic oscillations (BAO) as given by Sloan Digital Sky Survey (SDSS) \cite{BAO}. The BAO quantity enters only to impose constraints on $\Omega_{M}$. The possible departures of the DD relation  is based on two parametric representations for
$\eta(z)$ \cite{Holandaaa}, namely:
\\
\\
\hspace{1.0cm} I. $\eta (z) = 1 + \eta_{0} z$ \, \, and \, \, II. $\eta (z) = 1 + \eta_{0}z/(1+z)$.
\\
\\
\noindent  The first expression is a continuous and smooth linear one-parameter
expansion  while  the second one includes a possible epoch dependent
correction which avoids the divergence at extremely high z. 

The above parameterizations are clearly inspired on similar expressions for the $\omega(z)$-equation of state parameter in time-varying dark
energy models \cite{pad04,lind,cunha}. In the limit of extremely low redshifts ($z<<1$), we have $\eta = 1$ and $D_{L} = D_{A}$ as should be expected. Finally, we obtain $H_{0}$ constraints by marginalizing on $\eta_{0}$ and $\Omega_m$ parameters. Our results are compared with some recents $H_{0}$ measurements via galaxy clusters present in the literature  which took the DD relation as valid\cite{Boname06,Kobayashi96,reese02,cml,hcl}. 
    
\section{Galaxy Clusters Sample}

In the past few years the {\it Chandra } and {\it XMM}
observations of galaxy clusters  have shown that in general clusters
exhibit elliptical surface brightness maps, and so can not be
spherically symmetric. For this reason, De Filippis et al. 
\cite{DeFilippis05} reanalyzed and derived, using an isothermal
elliptical 2-Dimensional $\beta$-model to describe the clusters,
${{D}}_A$ measurements for 25 clusters from two previous
compilations: one set of data compiled by Reese {\it{et al.}}
\cite{reese02}, composed by a selection of 18 galaxy clusters
distributed over the redshift interval $0.14 < z < 0.8$  for which
high S/N detections of SZE, high S/N X-ray imaging and electron
temperatures were available, and the sample of Mason {\it{et al.}}
\cite{10}, which has 7 clusters from the X-ray limited flux
sample of Ebeling {\it{et al.}} (1996) \cite{Ebeling}. These two
previous compilations used a spherical isothermal $\beta$ model to
describe the clusters geometry. 

{ As
discussed by De Filippis et al. (2005) the choice of circular rather
than elliptical $\beta$ model does not afect the resulting of the
central surface brightness or Sunyaev-Zeldovich decrement, the slope
$\beta$ differs slightly between these models, however,
significantly different values for core radius are obtained. The
result was that the core radius of the elliptical $\beta$-model is
bigger than one of the spherical $\beta$ model (see Fig. 1 in their
paper). In first approximation it was found that $\theta_{ell} =
\frac{2e_{proj}}{1+e_{proj}}\theta_{circ}$, where $e_{proj}$ is the
axial ratio of the major to the minor axes of the projected
isophotes. Moreover, a mixture of simulations and studies of nearby clusters suggests a 10\% effect on the Hubble parameter  due to departures from isothermality. An extensive discussion of statistical and systematic errors of the SZE/X-ray technique can be found in references \cite{Boname06,reese02}.}

\section{Deviations in Cosmic Distance Duality and $H_{0}$ constraints from Galaxy Clusters}

In the last decade, some authors have used samples of ADD of galaxy clusters obtained via SZE/X-ray technique to constrain the $H_{0}$ value. The basic procedure has been to perform a qui-square statistics by considering the DD relation as valid and  use the theoretical expression for $D_{A}$ in the FLRW framework as, for instance, in a flat $\Lambda$CDM model. In this case the theoretical expression for $D_{A}$ is given by (c=1)\cite{lim03}  

\begin{equation}
{{D}}_{A}(z;H_0,\Omega_m) = \frac{H_{0}^{-1}}{(1 +
z)}\int_{o}^{z}\frac{dz'}{{{E}}(z';\Omega_m)}\,
\label{eq1}
\end{equation}
where $\Omega_m$ is the matter density parameter and the dimensionless
function ${{E}}(z';\Omega_m)$ is given by
\begin{equation}
{{E}} = \left[\Omega_m(1 + z')^{3} + (1 -\Omega_m)\right]^{1/2}.
\label{eq2}
\end{equation}
Some authors have fixed $\Omega_m$ \cite{Kobayashi96} while other ones have perfomed a joint analysis with other cosmological quantities to break the degeneracy on this parameter \cite{cml,hcl}.

However, if the DD relation does not hold, the measurements of X-ray and SZE give $D^{SZE/X-ray}_{A}(z)=D_{A}(z)\eta^{2}(z)$ \cite{uzan}. In this way, a more robust analysis should be performed by marginalizing on possible deviations of the DD relation (in our case, on $\eta_0$). So, in recent papers \cite{Boname06,Kobayashi96,reese02,cml,hcl}, the quantity which is being measured via clusters in the flat $\Lambda$CDM framework is 
\begin{equation}
{{D}}_{A}(z;H^{*}_0,\Omega_m) = \frac{1}{(1 +
z)H_{0}^{*}}\int_{o}^{z}\frac{dz'}{{{E}}(z';\Omega_m)}\,,
\label{eq5}
\end{equation}
where $H_{0}^{*}$ is $H_{0}^{*}=H_{0}/\eta^{2}$ if one takes the $\eta$ parameter as a constant. In our case, the $\eta$ can be a function of  $z$ and its influence on $H_0$ constraints should be take into account in statistical  analysis (see Sec. 4).

\begin{figure}[t]
\centerline{\epsfig{figure=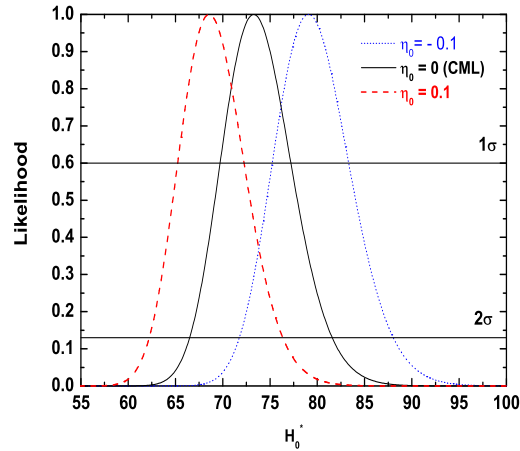,width=3.0truein,height=2.8truein,angle=0}}
\vspace*{8pt}
\caption{Likelihoods for the $H^{*}_{0}$ using differents $\eta_0$ values $(\eta_0 = -0.1, 0, 0.1)$ in linear $\eta(z)$ function  by using 25 angular diameter distance from galaxy clusters.Our analysis is based on a flat $\Lambda$CDM model and we fix $\Omega_m = 0.27$. The $\eta_0= 0$ (CML) corresponds to a previous work (see text).}
\end{figure}

Let us assume some $\eta_{0}$ values $(\eta_{0} = -0.1, 0, 0.1)$ in the linear parametrization to test the sensitivity of $H_{0}$ constraints obtained via SZE/X-ray technique when there is a violation of the DD relation but it is not considered in the statistical analysys. For now, the $\Omega_{M}$ is fixed in 0.27 value. We plot in the fig. (1) the likelihoods to $H_{0}^{*}$ obtained by using the 25 galaxy clusters from De Filippis {\it et al.}  sample \cite{DeFilippis05} for the different $\eta_0$ values. As one may see  if $\eta_{0}=0$, it follows that $D_A^{SZE/X-ray}(z)=D_{A}(z)$ and $H^{*}_{0}$ gives the expected Hubble parameter value, such as $H^{*}_{0} = H_{0} = 73.3$ km/s/Mpc (see Ref. \cite{cml} (CML)). However, if $\eta_{0} \neq 0$, and this fact is not taken into account, there is a remarkable difference between $H^{*}_{0}$ and $H_{0}$. In this way, if $\eta_{0} > 0$ it is direct that $D^{SZE/X-ray}_{A}(z)>D_{A}(z)$ and $H^{*}_{0} = 68$ km/s/Mpc $<H_{0}$. On the other hand, if $\eta_{0} < 0$ it follows that $D^{SZE/X-ray}_{A}(z)<D_{A}(z)$ and $H^{*}_{0} = 79$ km/s/Mpc $>H_{0}$.  So, a slight violation of the DD relation ($\eta_{0}=-0.1$ and $\eta_{0}=0.1$, which correspond to a  $5\%$ violation  for $z=0.5$, for instance)   furnishes   $H_{0}$ estimates mutually compatible only in $\approx 1.5\sigma$. Naturally, possible departures of the DD relation will be negligible  if only galaxy clusters in low redshifts ($z<0.1$) are used to constrain $H_0$.

\begin{figure*}[t]\label{Fig2}
\centerline{
\epsfig{figure=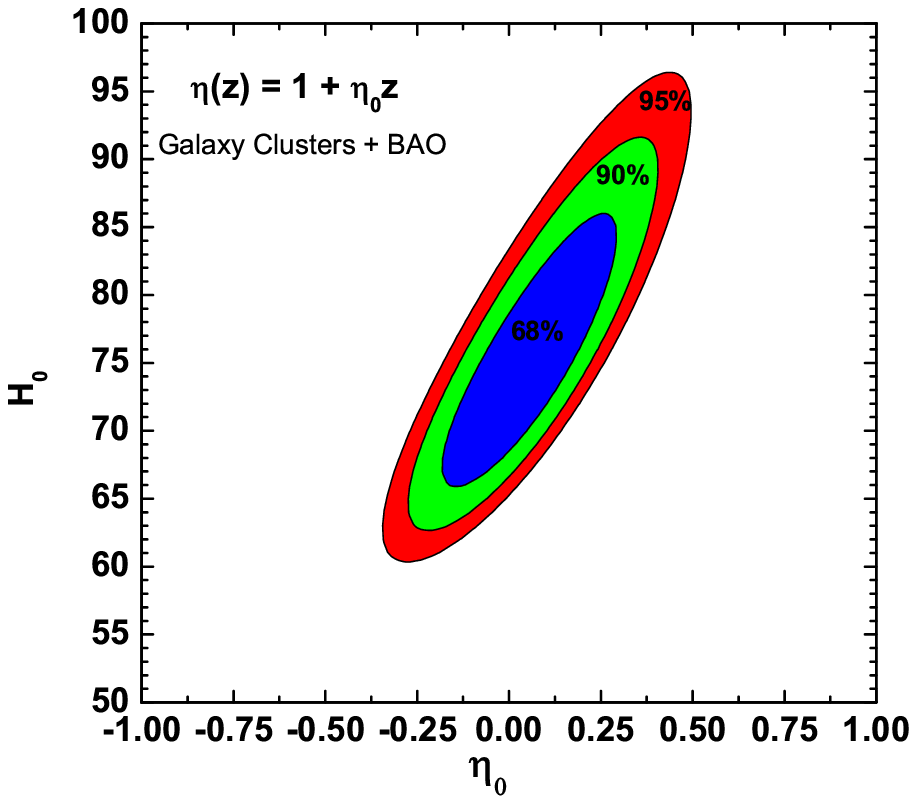,width=2.8truein,height=2.8truein}
\epsfig{figure=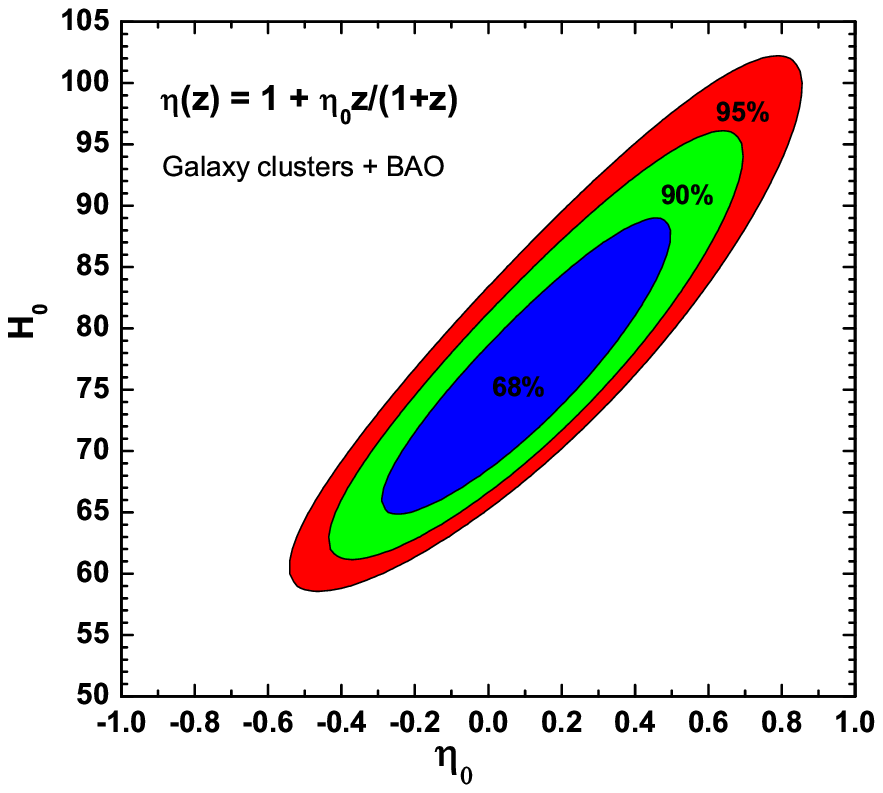,width=2.8truein,height=2.8truein}
 \hskip
0.1in} \caption{{ a)}Contours in the
$( H_{0},\eta_{0})$ plane using 25 angular diameter distances + BAO measurement  for the linear $\eta(z)$ function. {{ b)}Contours in the
$( H_{0},\eta_{0})$ plane using 25 angular diameter distances + BAO measurement for the non linear $\eta(z)$ function. In both case we have marginalized over $\Omega_{m}$.}
}
\end{figure*}

\section{Constraining $H_{0}$ and $\eta_{0}$ with X-ray/SZE Technique + BAO }

In this section we use the De Filippis {\it et al.} sample  \cite{DeFilippis05}  to impose constraints on $H_{0}$ and $\eta_{0}$ in both parametrization. Unlike previous works, we minimize the $\chi^2$ function that takes into account possible deviations of the DD relation ($\eta^{2}D_A(z)=D^{SZE/X-ray}_A $), given by

\begin{equation}
\chi^2(z|\mathbf{p}) = \sum_i { (\eta(z)^{2}{{D}}_A(z_i; \mathbf{p})-
{{D}}_{Ao,i})^2 \over \sigma^{2}_{{{D}}_{Ao,i}} } + \frac{(A-0.469)^{2}}{0.017^{2}},
\end{equation}
where ${{D}}_{Ao,i}$ is the observational ADD,
$\sigma_{{{D}}_{Ao,i}}$ is the uncertainty in the individual
distance, $\eta(z_i)$ is given by parametrizations linear and non linear and $A$ is the BAO measurement, namely \cite{BAO}
\begin{eqnarray}
 {{A}} \equiv {\Omega_{\rm{m}}^{1/2} \over
 {{{H}}(z_{\rm{*}})}^{1/3}}\left[\frac{1}{z_{\rm{*}}}
 \Gamma(z_*)\right]^{2/3}  = 0.469 \pm 0.017, 
\end{eqnarray}
where $z_{\rm{*}} = 0.35$ is the redshift at which the acoustic
scale has been measured and $\Gamma(z_*)$ is the dimensionless
comoving distance to $z_*$. Note that the above quantity is independent of the Hubble constant,
and, as such, the BAO signature alone constrains only the $\Omega_m$
parameter. The complete set of parameters is given by
$\mathbf{p} \equiv (H_0, \Omega_{m})$ and  $\eta_{0}$. For the galaxy clusters sample the common statistical contributions 
are: SZE point sources $\pm 8$\%, X-ray background $\pm 2$\%, 
Galactic N$_{H}$ $\leq \pm 1\%$, $\pm 10$\% isothermality, $\pm 8$\% kinetic SZ and for CMB anisotropy $\leq \pm 2\%$ \cite{reese02}.

\begin{figure}[t!]
\centerline{
\psfig{figure=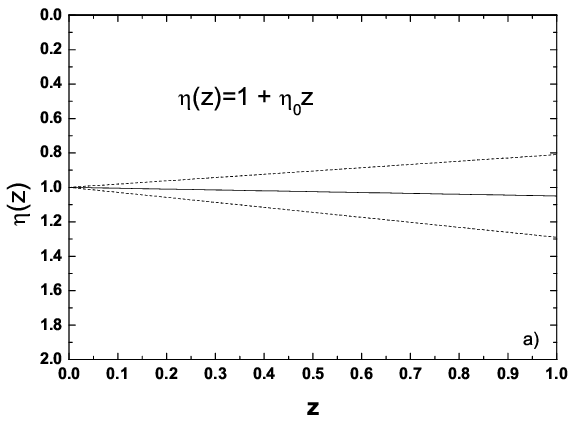,width=2.8truein,height=2.8truein,angle=0}\psfig{figure=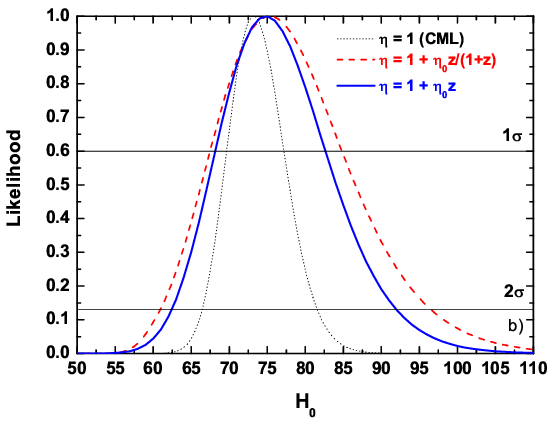,width=2.8truein,height=2.8truein,angle=0}
} 
\caption{ a) The $\eta(z)$ function for the linear parameterization. b)Likelihoods for the $H_{0}$ using 25 angular diameter distances and BAO measurement. Here we marginalized over $\Omega_{m}$ and $\eta_{0}$ for both $\eta (z)$ functions. We also plot the $\eta = 1$  case  in order to compare the likelihoods.} 
\end{figure}

In Figs. (2a) and (2b) we plot the contour regions ($H_0$, $\eta_0$) for each parametrization with ADD of galaxy clusters and BAO by marginalizing on $\Omega_m$. We obtain a strong correlation involving $h$ and  $\eta_0$. Our results are $\eta_{0} = 0.05^{+ 0.24}_{-
0.22}$ and $H_{0}= 75^{+ 11}_{-9}$ km/s/Mpc ($\chi^2 = 28.02$) in $68\%$ c.l. and  $\eta_{0} = 0.07^{+ 0.43}_{- 0.37}$ and  $H_{0}= 75^{+ 12}_{-10}$ km/s/Mpc ($\chi^2 = 28.05$) in $68\%$ c.l., without systematic errors, to the linear and non linear $\eta(z)$ functions, respectively. Although the confidence contours are large, we can see  that the ADD of the De Filippis {\it et al.}  sample\cite{DeFilippis05} is in agreement with no DD relation violation, $\eta_{0} = 0$, in $1\sigma$ and $H_{0}$ value is in perfect agreement with recent independent estimates  to  $H_{0}$, $H_{0} =74.2 \pm 3.6$ km/s/Mpc, a $4.8$\% uncertainty including both statistical and systematic errors \cite{riess}.{ In Fig. 3(a) we plot the $\eta(z)$ function to linear parameterization by using the best fit value and the respective $1\sigma$ error. As one may see, the standard result, $\eta(z)=1$, is compatible at $1\sigma$ c.l..}
 
\section{Constraining $H_{0}$ Regardless of the Distance Duality Validity}

Finally, by using the De Filippis {\it et al.} sample \cite{DeFilippis05}, constraints on $H_{0}$ regardless of the DD relation validity are obtained. In this point, we marginalized on $\Omega_{m}$ and the $\eta_{0}$ parameter in our likelihood. Our results are plotted in Fig. (3b). The constraints on $H_0$ are $H_{0}= 75^{+ 7}_{-7}$ km/s/Mpc  and $H_{0}= 75^{+ 10}_{-7}$ km/s/Mpc in $68\%$ c.l. for the linear and non linear parametrization, respectively. We also plot the likelihoood for $H_{0}$ in the $\eta_{0} = 0$ case \cite{cml} in order to compare the constraints. For this case $H_{0}= 73.3^{+ 4.5}_{-4.5}$ in $68\%$ c.l.. All values presented here are without systematic errors. So, the $H_{0}$ values obtained in our analysis are in excellent agreement with each other for the different $\eta$ functions (I and II) and are compatible in $1\sigma$ with the $H_0$ value from $\eta_{0} = 0$ case. In this way, our results proves the robustness of the recent $H_0$ measurements from X-ray and SZE clusters observations \cite{Boname06,Kobayashi96,reese02,cml,hcl}. 

In forthcoming SZE surveys, when a larger number of ADD  of galaxy clusters with lower systematics errors spread in the redshifts  will be available, we will be able to impose tighter constrains on $H_{0}$ and $\eta_{0}$ parameters.

\section{Conclusions}\label{sec:Conclusions}

In this work, we  explored constraints on $H_{0}$ from X-ray and SZE observations by assuming that a more general expression for the distance duality relation, such as $D_{L}(L)(1+z)^{-2}/D_{A}(z) = \eta(z)$, is valid. We show that if the distance duality relation does not hold, the constraints on the Hubble parameter in the FLRW framework via X-ray and SZE measurements actually furnishe an $H_{0}^{*}$, where $H_{0}^{*}$ is $H_{0}^{*}=H_{0}/\eta^{2}$. The $\eta(z)$ parameter was parametrized in two distinct forms, $\eta = 1+\eta_{0}z$ and $\eta = 1+\eta_{0}z/(1+z)$, thereby recovering the equality between distances only for very low redshifts. 

By using 25 angular diameter distances from galaxy clusters of the De Filippis {\it et al.} sample \cite{DeFilippis05} we  performed constraints on the Hubble and $\eta_{0}$ parameters simultaneously in a flat $\Lambda$CDM model. The  density matter parameter $\Omega_M$ was constrained  by adding the baryon acoustic oscillations (BAO) as given by Sloan Digital Sky Survey. By marginalizing on  $\Omega_M$ we obtain $\eta_{0} = 0.05^{+ 0.24}_{-
0.22}$ and $H_{0}= 75^{+ 11}_{-9}$ km/s/Mpc ($\chi^2 = 28.02$) in $68\%$ c.l. and $\eta_{0} = 0.07^{+ 0.43}_{- 0.37}$ and  $H_{0}= 75^{+ 12}_{-10}$ km/s/Mpc ($\chi^2 = 28.05$) in $68\%$ c.l., for linear and non linear $\eta(z)$ functions. Our results were consistent with a recent paper (see Ref. \refcite{Holandaaa}) where the De Filippis {\it{et al.}} sample was found to be in agreement with
the DD relation in the $\Lambda$CDM model
(WMAP7) \cite{komatsu} framework.

Finally, in order to test the robustness of the recent $H_{0}$ estimates by using X-ray and SZE technique we put constraints on the Hubble parameter by  marginalizing on $\Omega_{m}$ and  $\eta_{0}$ parameters. We obtained  $H_{0}= 75^{+ 7}_{-7}$ km/s/Mpc  in $68\%$ c.l. and  $H_{0}= 75^{+ 10}_{-7}$ km/s/Mpc in $68\%$ c.l., for linear and non linear parametrization, respectively.  For $\eta_{0} = 0$, $H_{0}= 73.3^{+ 4.5}_{-4.5}$ in $68\%$ c.l.. All values presented here are without systematic errors. So, the limits on $H_{0}$ in our analysis are in perfect agreement with recent independent estimates and with non violation of the distance duality relation. 

In the near future, when more
and larger samples with smaller statistical and systematic uncertainties become available, the method proposed here will be able
to improve the simultaneous limits on the Hubble and $\eta_{0}$ parameters.

\end{document}